\documentclass[useAMS,usenatbib]{mn2e}
\usepackage{graphicx}

\usepackage{epsfig}
\usepackage{epstopdf}
\usepackage{lscape} 

\def\lesssim{\mathrel{\hbox{\rlap{\hbox{\lower4pt\hbox{$\sim$}}}\hbox{$<$}}}}
\def\gtrsim{\mathrel{\hbox{\rlap{\hbox{\lower4pt\hbox{$\sim$}}}\hbox{$>$}}}}
\title[Interstellar Plasma Scattering]{
Pulsar scintillations from corrugated reconnection sheets in the ISM
}
\author[Pen and Levin]{Ue-Li
  Pen,$^{1}$\thanks{E-mail:\ pen@cita.utoronto.ca}
Yuri Levin,$^2$\thanks{E-mail:\ yuri.levin@monash.edu}\\
  $^1$ Canadian Institute for Theoretical Astrophysics, University of Toronto, M5S 3H8 Ontario, Canada \\
$^2$ Monash University, Australia
}
\begin{document}

\date{\today}

\pagerange{\pageref{firstpage}--\pageref{lastpage}} 
\pubyear{2012}

\maketitle
\label{firstpage}
\begin{abstract}

We show that surface waves along interstellar current sheets closely
aligned with the line of sight lead to pulsar scintillation properties
consistent with those observed.  This mechanism naturally produces the
length and density scales of the ISM scattering lenses that are
required to explain the magnitude and dynamical spectrum of the
scintillations.  In this picture, the parts of warm ionized
interstellar medium that are responsible for the scintillations are
relatively quiescent, with scintillation and scattering resulting from
weak waves propagating along magnetic domain boundary current sheets,
which are both expected from helicity conservation and have been
observed in numerical simulations.  The model statistically predicts
the spacing and amplitudes of inverted parabolic arcs seen in
Fourier-transformed dynamical spectra of strongly scintillating
pulsars with only 3 parameters. Multi-frequency, multi-epoch low
frequency VLBI observations can quantitatively test this picture.  If
successful, in addition to mapping the ISM, this may open the door to
precise nanoarcsecond pulsar astrometry, distance measurements, and
emission studies using these 10AU interferometers in the sky.

\end{abstract}

\newcommand{\be}{\begin{eqnarray}}
\newcommand{\ee}{\end{eqnarray}}
\newcommand{\beq}{\begin{equation}}
\newcommand{\eeq}{\end{equation}}

\section{Introduction: observations and theoretical challenges}

Compact radio sources provide a precision probe of the ionized
interstellar medium (IISM).  The propagation speed of radio waves
depends on the density of free electrons, and therefore the spatial
inhomogeneity of the IISM may result in refractive and diffractive
abberation and scattering. This causes scintillation
(time-variability) of the compact radio-sources
\citep{1968Natur.218..920S,1986ApJ...301L..53B,2006ApJS..165..439R}.
Observations of the pulsar scintillations are particularly interesting
for infering the IISM properties, due to the brightness of many
pulsars that have been observed for other purposes over long time
intervals.

However, despite of considerable effort, the small-scale structure of
the IISM has remained enigmatic. At a first glance, one expects the
scattering to be caused by density inhomogeneities produced from
turbulent motions of the IISM. However, this picture alone
is unable to explain the last decade of the pulsar scintillation data.
There, a major observational progress of the ISM scintillations has
been achieved through the \citet{2001ApJ...549L..97S} detection of
inverted parabolic arc structures in the Fourier-transformed dynamical
spectra of some
strongly scintillating nearby pulsars at some epochs. These inverted
parabolic arclets imply that 
for these pulsars the radio-wave scattering occurs mostly within one
or a few thin screens
\citep{2004MNRAS.354...43W,2006ApJ...637..346C,2008MNRAS.388.1214W}.
Moreover, the multiple ``inverted parabolae''
\citep{2005ApJ...619L.171H} show that the scattering inside the screen
is strongly inhomogeneous and occurs in localized clumps. The latter
was recently confirmed by \cite{2010ApJ...708..232B} who have constructed
the scintillation derived scattering image of PSR B0834+06 from VLBI
data. They have found that not only  
the scattering image was clumpy but that the clumps lined up along a
thin line.  Refractive effects are usually invoked to explain these phenomena
\citep{2006ApJ...637..346C}.  The physical conditions to create these
scattering angles are mysterious, and have been suggested to hint at
new physical
phenomena\citep{2007ASPC..365..299W,2001Ap&SS.278..149W,2013PhLB..727..357P}. 
\cite{2006ApJ...640L.159G} showed that
the interference of images in refractive lensing also results in
scintillation which resembles purely diffractive processes.

We use the term ``scattering'' to refer to the general phenomenon of
change of light propagation direction, for which one can consider
different physical mechanisms, including diffraction and refraction.


In this paper, we explore the quantitative consequences of the
refractive scintillation picture. Since refractive effects are
required in any case, we explore the possibility if these alone can
predictively explain the observations, and how this can be tested in
the future.  We study a scenario in which the scattering is produced
by a refractive structure, with turbulence playing a secondary role.
Radio-wave scattering by non-turbulent large-scale refractive
structures has been previously considered by
\cite{1987Natur.328..324R}, mostly in the context of the so-called
extreme-scattering events observed by \cite{1987Natur.326..675F}.
Recently, \cite{2006ApJ...640L.159G} proposed that the image of the
SgrA* radio-source is strongly scattered by several reconnection
sheets that are closely aligned with the line of sight to SgrA*.  In
this paper, we develop further the Goldreich \& Sridhar's (2006) idea
[see also \cite{2012MNRAS.421L.132P}] and apply it to construct a
quantitative picture of pulsar scintillations. Namely, we consider a
scenario where the pulsar radio-wave scattering occurs due to several
{\it weakly corrugated} reconnection sheets that are closely aligned
with the line of sight to the pulsar; see figure \ref{fig:sheetgeom}.
We show that this scenario provides explanations for previously
unexplained features of the scintillations: (1) the ``scattering
screens'' are simply effective descriptions of such sheets; their
location is marked approximately by the sheets' intersections with the
line of sight, (2) ``the scattering clumps'' correspond to those parts
of sheet folds where the sheet is parallel to the line of sight; the
strength of the scattering follows a strongly non-Gaussian
distribution, even though the corrugation itself is assumed to be a
realization of a Gaussian distribution, and (3) the strong
non-isotropy of the Brisken et al.~2010 image is a consequence of the
sheet's inclination, with the clump locations aligned along the
direction perpendicular to sheet's line of nodes. The width of the
image is given by the inclination angle of the sheet. (4) the high
degree of alignment is a selection effect, as described by
\cite{2006ApJ...640L.159G}: all sheets aligned by less than the
critical angle do not project into fold
singularities
\citep{citeulike:4872486}
and thus do not
contribute to strong lensing. The plan of the paper is as follows: in
the next section we briefly describe the origin of the reconnection
sheets, in section 3 we derive the model for the fold statistics, in
section 4 we derive the lensing by the corrugated sheet, and in
section 5 we compute the Fourier-transformed dynamical spectrum and
demonstrate the emergence of inverted parabolic arclets. In section 6
we discuss physical implications, section 7 future speculation, and conclude
in section 8..

\section{Astrophysical Picture}

\subsection{Two Regimes of Lensing: Diffractive vs Refractive}

Two regimes to generate pulsar scintillation have been considered. 
Traditionally, scintillation has been treated as the scattering of
radio waves by a volume filling random field, produced by a turbulent
cascade \citep{1990ARA&A..28..561R}.  This process produces structures
on a large range of scales.  In this paper, we will classify the
lensing into two distinct regimes:

We define the diffractive limit of wave scattering
when the  angle is determined by
the ratio of the wavelength $\lambda$ of the radio-waves, and the
characteristic spatial scale $D$ of the scattering structure:
$\theta\sim\lambda/D$.  The brightness of the scattered image is
determined by the amplitude of the wavefront modulation caused by the
scattering structure.  To explain the observed angles in the range
$1-100$ mas at wavelength $\sim 1$m, requires structures in the ISM on
transverse scales of order $10^{6-8}$m. 
This imposes unexpected properties on the
IISM, since this scale much smaller than the coloumb mean free path of
protons and thus compressive perturbations of this scale are
overdamped and decay exponentially on the sound-crossing time-scale,
which is hours to days.
If one still assumes that somehow these perturbations are created and
maintained, then the angular image of a pulsar, and therefore its
dynamic spectrum, are superpositions of thousands of
$\hbox{AU}/10^8\hbox{m}$ weak structures (for a 1-d scattering image),
and are expected to be roughly Gaussian.  The length scale is given by
the path length difference, which in turn is inferred from the inverse
scintillation correlation frequency.  The asymmetric parabolically
structured 2-D power spectrum of the dynamic spectrum containing
inverted parabolic arcs, and the VLBI
image of the scattering disk consisting of several prominent clumps,
are inconsistent with such a picture.  The number of $10^{6-8}$m
eddies along the line of sight can be large, possibly $10^{13}$ for a
pulsar at a distance of $\sim$ kpc.  The total scattered power comes
from the cumulative projected variation in refractive index, which
grows as the square root of the number of eddies.  Each eddie only
needs to change the refractive index by a part in a million of the
cumulative change, so a refractive index variation of a part in $\sim
10^{12}$ could account for the strong scintillation.  But the
superposition of scattering from such a large number of eddies would
surely look very Gaussian by the central limit theorem.  A requirement
for the sum of $10^{13}$ contributions to appear intermittent in
projection would lead to unphysically overpressurized eddies.

A second mechanism is due to refractive lensing.  For structures
larger than the Fresnel scale 
\begin{equation}r_F \equiv \sqrt{\lambda L} \sim 0.03
AU \left(\frac{\lambda}{{\rm m}}\right) \left(\frac{L}{{\rm kpc}}\right)
\end{equation}
deflection of light rays are described by geometric optics.  In this
regime, points of stationary phase in phase optics correspond to local
extrema of Fermat's principle, leading to multiple images of a
source.  The lens itself need not have structures on the diffractive
scale, it is sufficient for the phase to linearly change by $2\pi$
on a scale $D=\lambda/\theta$.  
When all structures in the lens are larger than $r_F$, the
refractive images will still show interference with each other, which
resembles some of the properties of DISS.
We will call this the refractive scintillation limit.

To further simplify this picture, we consider
the bending angle as determined by Snell's law, i.e. the change in
refractive index and the angles of incidence.  The refractive index of
a plasma at frequency $\omega$ is $n=1/\sqrt{1-\omega_p^2/\omega^2}$,
with plasma frequency 
$\omega_{\rm p}=\sqrt{n_e e^2/\epsilon_{0}m_e}$.  Pulsar observations
are done at frequencies much higher than the plasma frequency, for
which we expand $n-1 \sim -\frac{\omega_p^2}{2 \omega^2} = -1.8\times
10^{-8} n_e$ at wavelengths of a meter.  The observed scattered images
at 20 mas require deflection angles of at least 40 mas, corresponding
to $n_e \sim 10/{\rm cm}^3$ neglecting geometric alignment factors.
However, the mean density of the IISM is determined to be of the order
$10^{-2}$, as measured from the pulsars' dispersion
measure\citep{2004hpa..book.....L}.  Therefore, the refractive picture
is also challenging to reconcile with the data, since the observed
scattering angles would na\"ively require fractional changes in free
electron density of $\sim 10^3$ which are difficult to understand or
confine. This contraint, however, is alleviated when one considers
refractive sheets that are closely aligned with the line of sight
\citep{2006ApJ...640L.159G}.  At this grazing incidence, expressed
as an angle $\alpha$ from the surface of the lens, the deflection angle
is amplified to $\Delta \theta = n_2/n_1/\alpha$, so a sufficiently small
$\alpha$ can result in a large deflection.

Historically, refractive lensing was used to interpret long time
variability, and diffraction for the minute time scale
effects. Recently, it was understood that the refractive images result
in an interference fringe pattern \citep{2004MNRAS.354...43W} which
have similar time and frequency scales for flux modulation as
diffractive effects.  The frequency and time scaling was historically
interpreted as related to an underlying stochastic diffractive process
driven by turbulence.  \cite{2006ApJ...640L.159G} showed that
refractive lensing by aligned sheets results in scintillation similar
to the diffractive picture.



\subsection{Physical origin of the reconnection sheets}

The interstellar medium is stirred on scales of parsecs by various
energetic processes, including supernovae, ionization fronts, spiral
density waves, and other phenomena.  These processes are generally
short lived, and we conjecture after the stirring, the warm medium
relaxes to an near-equilibrium configuration on small (several AU)
scales. Current numerical simulations of the supernova-driven
turbulence in the warm ISM of the Galaxy (e.g., Hill et al.~2011) do
not have the resolution to tell how realistic this assumption is.  In
the presence of helicity, the equilibrium magnetic fields are
configured as interlaced twisted tori, which are long lived
\citep{2004Natur.431..819B}.  It has been shown by \cite{2009arXiv0909.1815G}
that a ``generic magnetic equilibrium of an ideally conducting fluid
contains a volume-filling set of singular current layers.'' In this
picture, the magnetic fields are locally almost parallel, with
discontinuous interface regions, a bit like magnetic domains in a
ferromagnet. Singular current sheets have also been seen in the
simulations of \citet{2004PhRvL..92h4504S}.

At the boundary between between magnetic field configurations, current
sheets maintain the discontinuities.  Depending on the nature of
reconnection, these current sheet may be self-enforcing due to inflow
of fresh fields, maintaining a thickness potentially as thin as an
electron gyromagnetic radius.  At constant temperature,
the pressure equilibrium in the direction
perpendicular to the sheet implies that the electron density inside the sheet is
enhanced by a factor $R$ given by
\begin{equation} 
R-1\sim {B_{\rm out}^2-B_{\rm in}^2\over 4\pi P_{\rm out}},
\label{ratio}
\end{equation}
where $B_{\rm in}$ and $B_{\rm out}$ is the magnetic field in and
outside the sheet, and $P_{\rm out}$ is the non-magnetic pressure
outside the sheet. The right-hand side of the above equation is
thought to be of order $1$ for the IISM, but may be significantly
larger if the IISM is magnetic-pressure dominated.

Depending on the ratio of heating time due to reconnection to cooling
time, the thin sheet may have an enhanced temperature, in which case
it can be underdense, with $R \rightarrow 0$ in the limit of a
strong entropy injection.

\begin{figure*}
\centerline{\epsfig{file=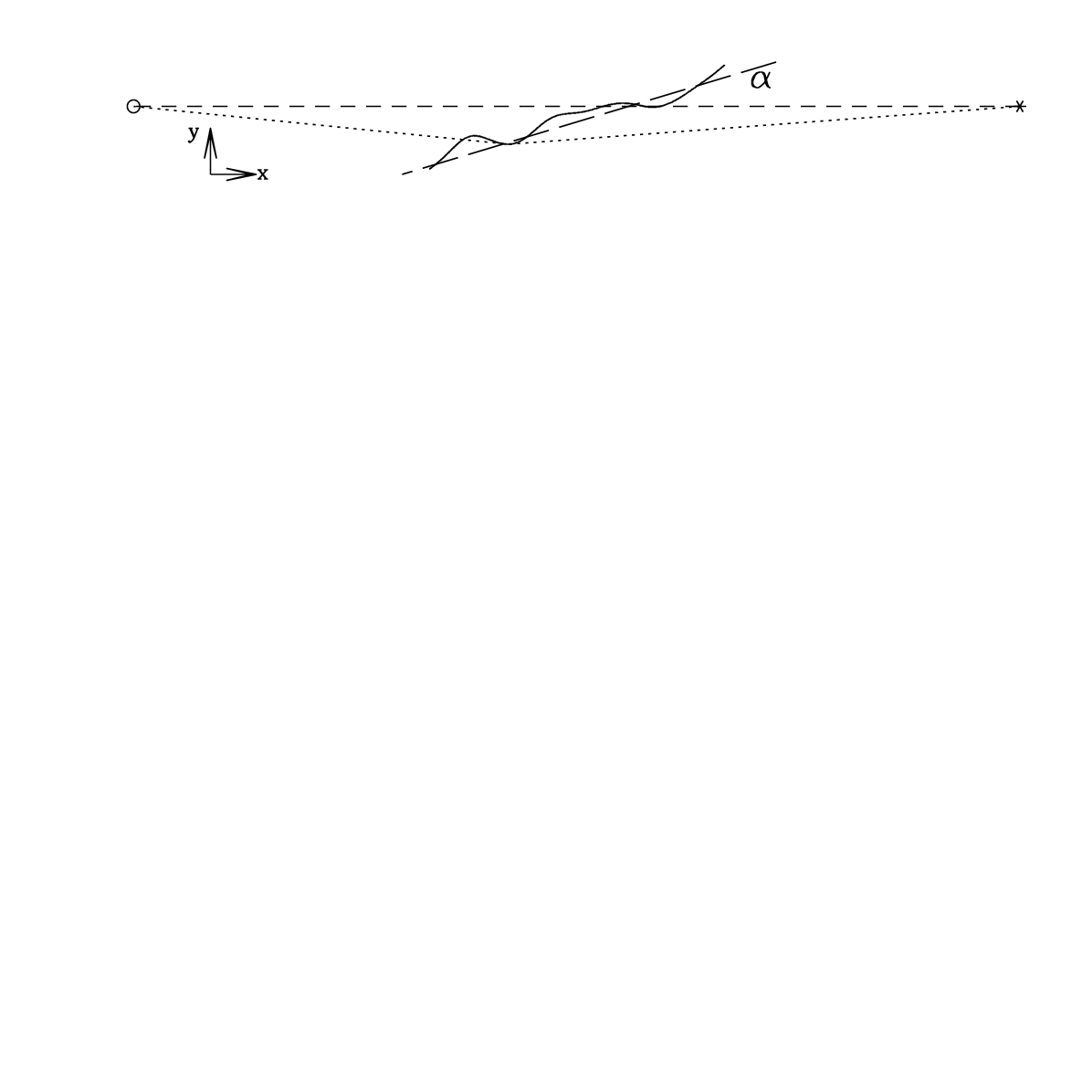,width=7.5in}}
\vspace{-6in}
\caption{lensing geometry.  The earth is at left, pulsar at right.
A section of the (crumpled) scattering sheet is shown in the middle as the solid
line.  It may be a plasma under- or overdensity of constant thickness.
The dashed line  
shows the  unperturbed light path.  The dotted line shows the path of 
an image
lensed by a fold singularity of the projected sheet.  The long dashed line
shows the position of the sheet in the absence of waves.
The inclination between this sheet and the line of sight is
$\alpha$, in this plot 1/8, chosen to exagerate the effect for
visual clarity, and
reduce the computational cost.  In practice, we envision values closer
to $10^{-3}$.
}
\label{fig:sheetgeom}
\end{figure*}

The lensing geometry is shown in Figure \ref{fig:sheetgeom}. A crucial
ingredient in our model is that the reconnection sheet is assumed to
be weakly corrugated. Let us assume that the corrugation pattern is
fixed, and vary the inclination angle. In this case, as can be seen
from figure \ref{fig:sheetgeom}, there is characteristic inclination
angle $\alpha$ at which the perturbations in the sheet can generate
fold singularities in the projected surface density.  This angle $\alpha$ is the
ratio of the wave peak displacement to its wavelength.  We envision
ranges of $\alpha \sim 10^{-3} - 10^{-2}$.  When the fold singularities are
present, they become effective refractive scattering centers for the
radio waves. The resulting lense features multiple images that closely
line up along the direction perpendicular to the line of nodes of the
scattering sheet.

\subsection{Surface Dynamics}

We assume that the current sheet is physically thin, $\lesssim $ AU.
Theoretically, the thickness of current sheets is not understood, so
we choose this scale to be thin enough to explain the smallest scale
observed structures.  On each side the magnetic field points in a
different direction.  The change in Alv\'enic properties gives rise to
surface waves \citep{1991SoPh..133..263J, 2009GApFD.103...89J}, whose
amplitude decays exponentially with the distance to the current sheet
and which are mathematically analogous to deep water ocean waves.  The
restoring force is due to the difference in magnetic field component
projected along the wave vector.  Like ocean waves, these waves
penetrate about a wavelength into each side.  We will be considering
wavelengths of hundreds to thousands of AU whose projected wavelength
is related to the observed lensing structures $\sim$several AU, so the
thickness of the current sheet itself is neglible as far as the
dynamics of the waves are concerned.  Seen in projection along the
aligned sheet, the projected wavelengths will be $\sim $ AU, reduced
by the alignment angle $\alpha$.

The displacements are transverse to the wave vector, and perpendicular
to the sheet.  While Alv\'enic in nature, the surface modes possess only
one polarization, unlike bulk waves.  We speculate
such modes to be long lived, since the single polarization nature
protects them from the normal MHD turbulent
cascade\citep{1997ApJ...485..680G}.  These waves resemble a flag
blowing in the wind.  Disturbances travelling along the sheet are
decoupled from bulk waves.  Being confined to a sheet, the amplitude
away from a source drops as $\propto 1/r$ instead of the normal
inverse square law.  The 2-D analogy to Olber's Paradox leads to flux
being dominated by the cumulative effect of far away sources, like
waves on an ocean beach.  We speculate potential sources to include
shock waves, spiral arms, stellar perturbations.
A amplitude of order $\alpha \lambda_{\rm
  wave}$, or about $10^{-3}$--$10^{-2}$ of the wavelength of the
surface wave, is sufficient to cause the sheet to appear folded in
projection; here $\lambda_{\rm wave}$ is the wavelength of the surface
wave.

\section{Fold Statistics}


We expect a minimum wavelength of these surface waves, which is
substantially larger than the thickness of the sheet.  Short
wavelength perturbations are not bound to the surface, and can
dissipate into the bulk.  The exact cutoff depends on unknown factors,
including the distance to the source, and vertical structure of the
current sheet.  Non-linear effects also cause short wavelength waves
to dissipate, just like sound waves in the air.  Primarily waves with
amplitude larger than $\alpha$ contribute to scattering.  The strength
of damping depends on the distance to the source in units of
wavelength, with shorter wavelengths being more damped.

\begin{figure}
\vspace{-0.8in}
\centerline{\epsfig{file=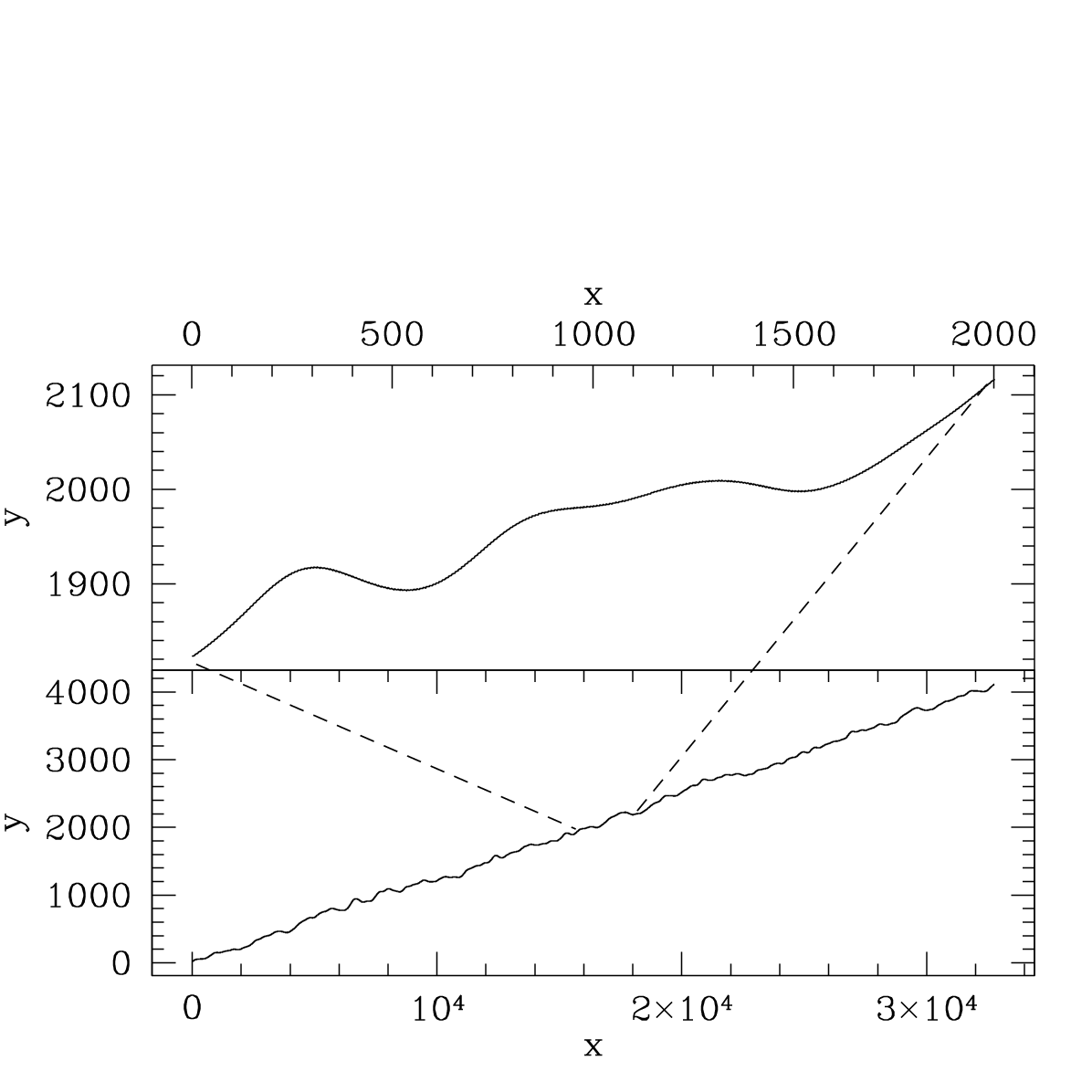,width=3in}}
\caption{Sheet with transverse perturbations.  The upper panel shows
  an zoomed version of a short section.  The $x$ axis is along the
  line of sight, the $y$ axis is transverse.  In a physical setting,
the grid units roughly
  map to A.U. for the $x-axis$, and deci A.U. for the $y-axis$.  The
  difference in conversion comes from the order of magnitude smaller
  inclination in the real system from the picture shown.}
\label{fig:sheet}
\end{figure}

We model the waves as a displacement function $\zeta(x)$
which is a Gaussian random field with a correlation function that is a
Gaussian, 
\begin{equation}
\xi(r)=\langle \zeta(x)
\zeta(x+r)\rangle=A^2\exp\left(-{r^2\over 2\sigma^2}\right),
\end{equation}
where $A$ is the mean amplitude
of the displacement and $\sigma$ is the surface-wave
coherence scale.  The coordinates are as denoted in Figure
\ref{fig:sheetgeom}: $x$ is the along the line of sight, which is
closely aligned to the long axis along the sheet, and we use the same
label for both. $y$ refers
to the position as seen on the sky.
We denote the projected coherence scale $\sigma_y
\equiv \alpha \sigma$.  Figure \ref{fig:sheet} shows a realization of a
sheet with a random fluctuations.  The displacement $\zeta$ is
measured perpendicular to the sheet, and thus neither $x$ nor $y$.
Since we are considering highly inclined sheets, $\zeta \sim y$ to
within $\alpha$, and we will use these two interchangably.
To reduce the computational cost,
we used an inclination slope
$\alpha=1/8$ instead of the physical range $\alpha \sim
10^{-3}-10^{-2}$, and a correlation length along the sheet of 350
units, 
which is $\sigma_y\sim 44$ grid units in projection.  The displacement
amplitude is chosen as $A=40$.  The dimensionless fluctuation
$\delta\equiv A/\sigma \sim 1$ is about unity, meaning a $1-\sigma$
fluctuation can result in a fold singularity, which we will discuss further
below.

\begin{figure}
\centerline{\epsfig{file=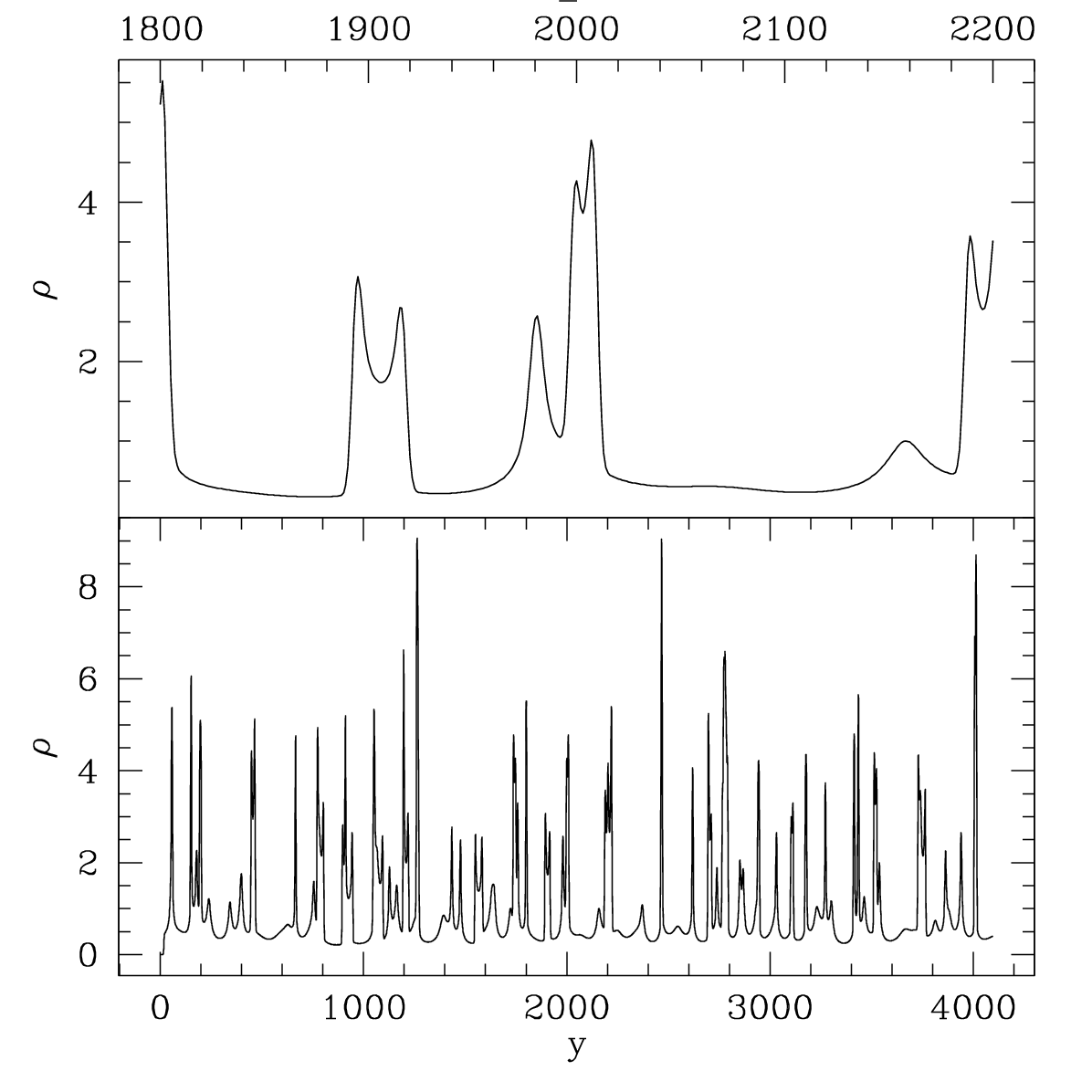,width=3in}}
\caption{projected density.  The upper panel is a zoom of the central 
  portion of the lower panel. Each time the sheet folds in projection,
we see a double peaked  structure in density.  The
characteristic separation
between the two peaks is a projection correlation length $\sigma_y$,
in this case 44 grid units.}
\label{fig:rho}
\end{figure}

The current sheet corresponds to a change in magnetic and thermal
pressure and thus has a refractive index different from the ambient
ISM. For Alv\'enic waves, we can treat the sheet as constant
thickness, and consider its convergence as is relevant for
refractive lensing.  In projection along the line of sight, the column
density of the sheet results in a highly non-Gaussian
distribution. Figure \ref{fig:rho} shows the column density
distribution in a simulation.  Folds occur when the gradient of the
displacement is equal to $-\alpha$, which as described above is
quantified by the fluctuation amplitude $\delta$.  The value chosen
here makes folds common, occuring roughly once per correlation length,
and yet make multiple folds overlapping in projection rare.  Each fold
results in two fold singularities, with characteric separation $\sigma_y$. This
is well understood from the theory of extrema of surface
waves\citep{1957RSPTA.249..321L}.

%
%

Equation (\ref{eqn:prho}) describes the 1 point PDF corresponding to
the spatial density distribution shown in  figure
\ref{fig:rho}.  The deflection angle is determined by the gradient of
the density, $\rho'$.


As in \cite{2012MNRAS.421L.132P}, we use the notation of gravitational
lensing\citep{1992grle.book.....S}.  The phase delay through the lens
is described by the lensing potential  
\begin{equation}
\psi\equiv  \frac{\Delta z \omega_p^2}{2 L\omega^2}
\end{equation}
where $\Delta z$ is the thickness of the sheet along the line of
sight.   $L$ is the distance to the pulsar. 
The convergence $\kappa \equiv \partial_\theta^2\psi/2$ is given by the
second angular derivative 
of the potential, and thereby the laplacian of the projected density
$\rho$.  The  
magnification gives the 
brightness of the image.  In 1-D it is $\mu=1/(1-2\kappa)$, derived
from the determinant of the amplification matrix.  A negative
amplification corresponds to a flipped image of odd parity.

As before, $\alpha$ is the angle between the screen and the line of
sight.  Then the convergence is $\kappa\propto 1/\alpha$ for
$\alpha\ll 1$, and $P_{\rm screen}(\kappa)\propto 1/\kappa^2$, where
$P_{screen}(\kappa)$ is the probability of a piece of screen to have
convergence of $\kappa$.  However, we are interested in the
probability density with respect to the impact parameter relative to
the line of sight, and not with respect to the location on the
screen. For nearly aligned screens, this is not the same thing. In
particular, part of the screen with low $\alpha$ occupies less of the
impact-parameter space than the part of the screen with the same area
but high $\alpha$. Thus,
\begin{equation}
P_{\rm impact parameter}(\rho)\propto 1/\rho^3
\label{eqn:prho}
\end{equation}
which is a strongly non-Gaussian distribution.

The regions near the locations where the screen is parallel to the
line-of-sight, called fold singularities, give rise to the localized
clumps in the pulsar's scattering image, which produce the inverted
parabolic arcs in pulsar secondary spectra.

\section{Lensing}

The lensing of this density sheet can be computed in analogy to 
 \cite{2012MNRAS.421L.132P,1998ApJ...496..253C}.

\begin{figure}
\centerline{\epsfig{file=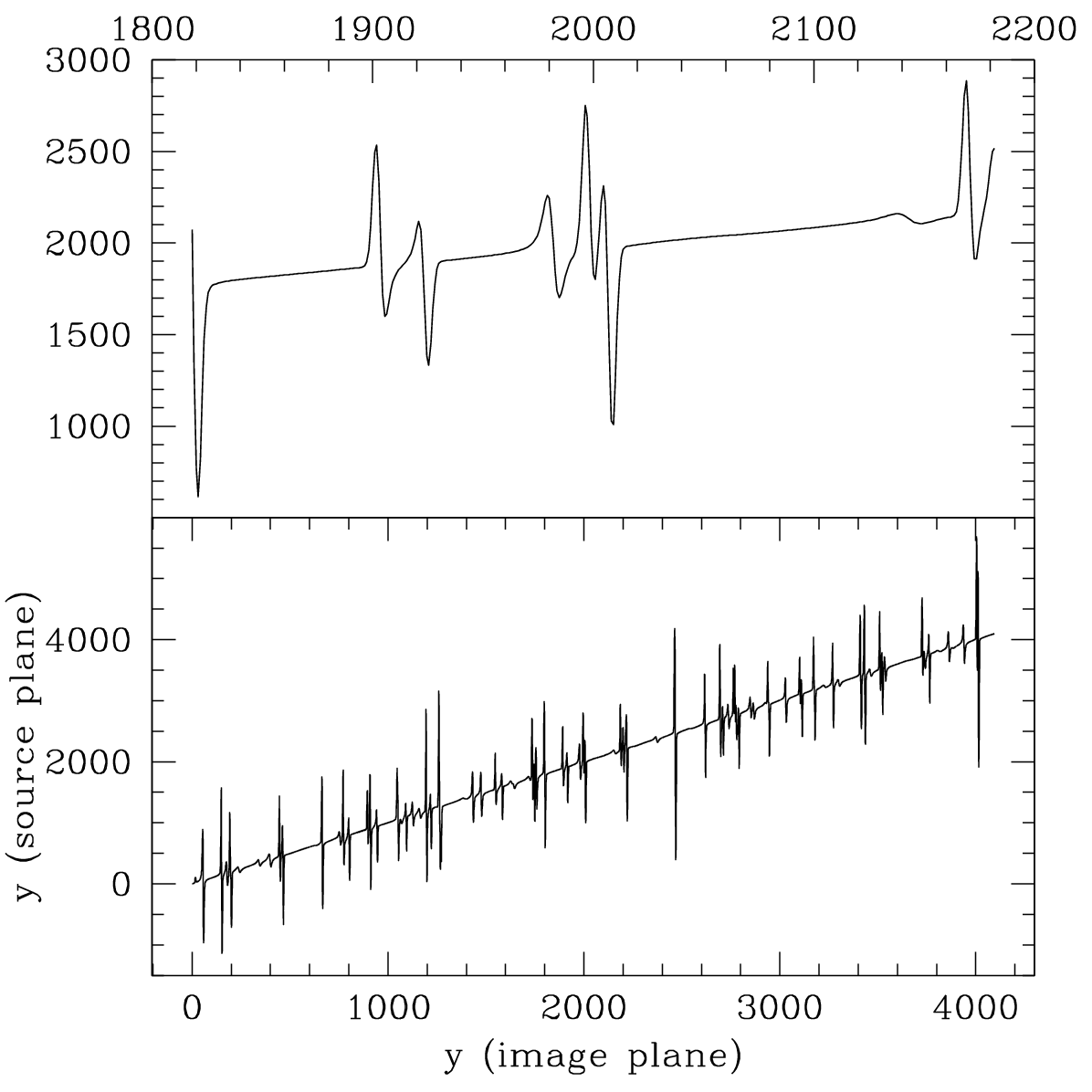,width=3in}}
\caption{deflection angle mapping. The horizontal axis is angle on the
sky, and the vertical axis is the intersection of this light ray on
the source plane.  Whenever multiple different directions on the sky
intersect on the same position in the source plane, multiple images
are formed, which form a coherent interference pattern. As before, the
upper panel is a zoom of the lower panel.}
\label{fig:dt}
\end{figure}

Given the projected column density distribution in Figure \ref{fig:rho}, and assuming that
the sheet has a fixed thickness along its normal, we
can compute the mapping of apparent angle on the sky to position in the
source (pulsar) plane.  The fold singularities in the projected density
distribution lead to large angle deflections, and multiple images,
whenever the sheets are aligned closely enough for the folds to form.
This explains why only a small fraction of the current
sheets contribute to scintillation. 

The system depends on the three dimensionless parameters: the
projected perturbation amplitude $\delta$, the
ratio $r$ of sheet thickness $z$ to the projected correlation length
$\sigma_y$, and the characteristic overdensity $C$ at a fold
singularity, described in more detail below..
The model predicts the number density of images and their fluxes as a
function of their angular separation from the line of sight (and
therefore as the function of time lag).  



There is a
dimensional scaling of time units, which is a function of pulsar
transverse velocity, projected screen size, and observing frequency.
This is generally parameterized as the DISS time scale, $t_{\rm DISS}
\sim (\lambda/D) (L/v)$ where $D$ is the size of the lensing region,
$L$ is the distance to the pulsar, $\lambda$ is the observing
wavelength, and $v$ is the pulsar transverse velocity.  Note that in
this model, the lensing is refractive, sharing the time scales from
diffractive models, but not the length scales.

\begin{figure}
\centerline{\epsfig{file=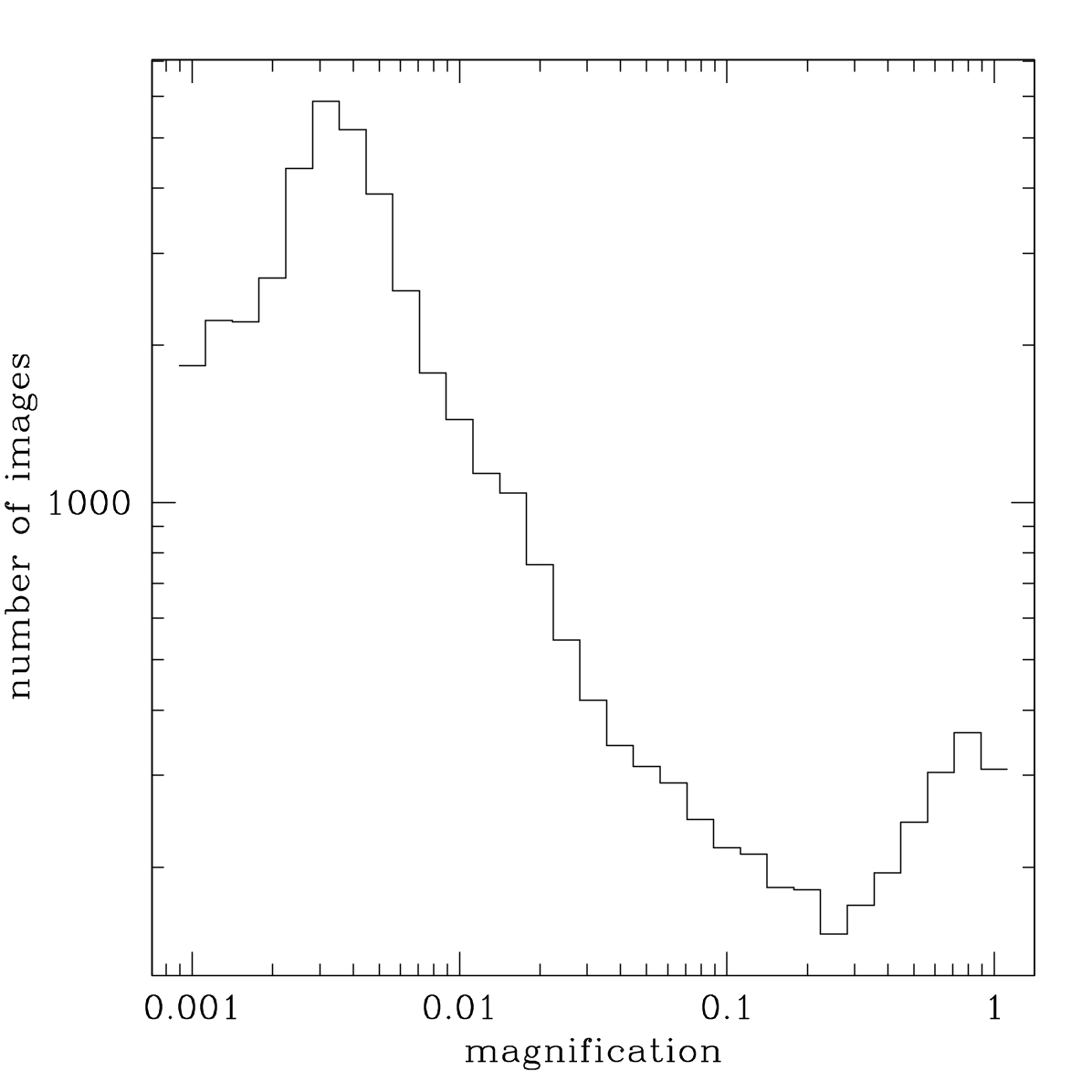,width=3in}}
\caption{PDF of image magnifications.  The peak near 1 are images at
  the unscattered positions, while the population on the left are
  lensed images.  The peak occurs at roughly $1/\kappa$.
}
\label{fig:mhist}
\end{figure}
\begin{figure}
\centerline{\epsfig{file=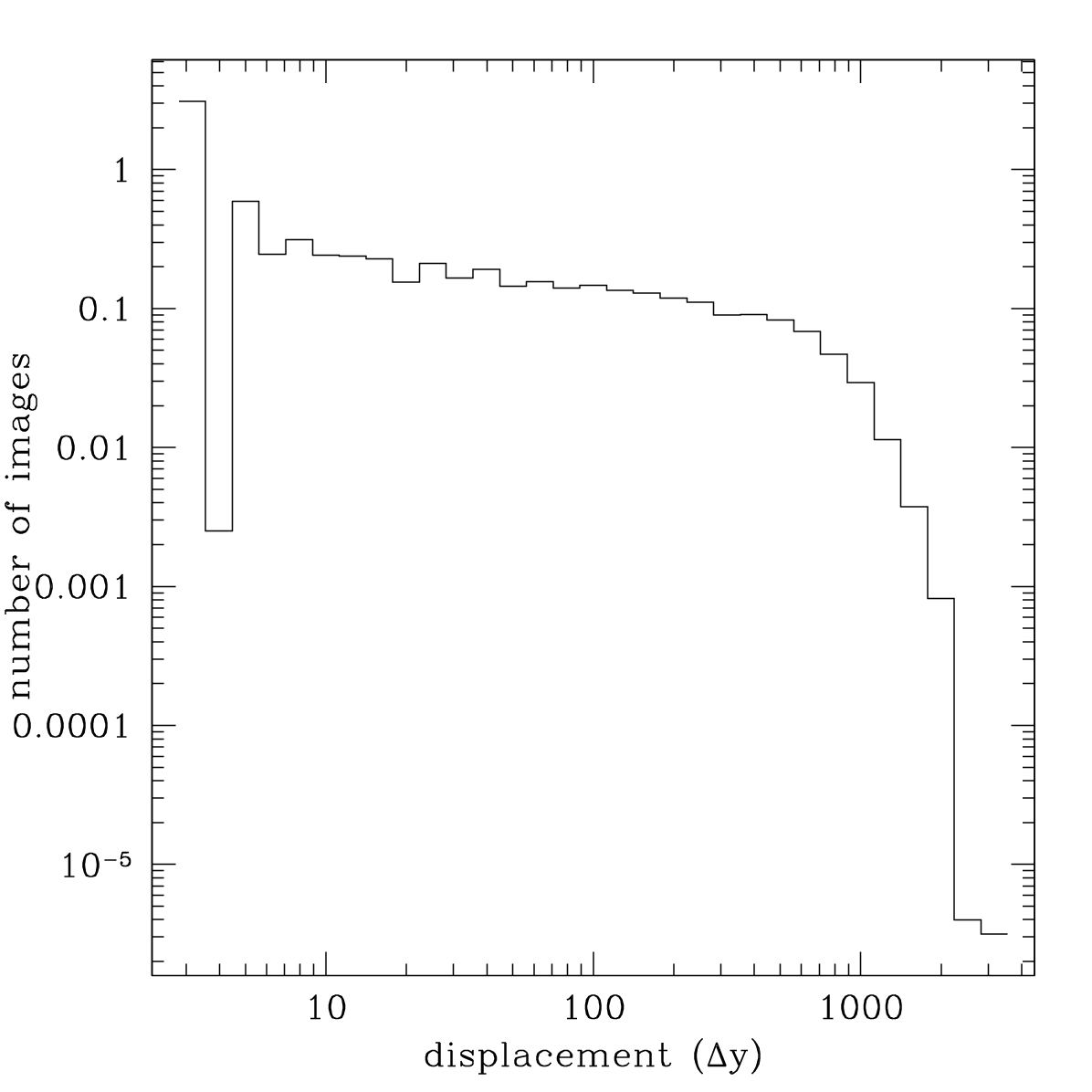,width=3in}}
\caption{PDF of image positions per logarithmic distance interval on
  the sky $y$.  The projected correlation length is $\sigma_y \sim
  44$.  The characteristic convergence $\kappa \sim 20$, leading to a
  cutoff near $\sigma_y \kappa \sim 1000$.
}
\label{fig:pos}
\end{figure}

We show a histogram of image magnifications in Figure \ref{fig:mhist},
which can be compared to holographic flux measurements
\citep{2008MNRAS.388.1214W}. The lensed images correspond to the
inverted arclets seen in secondary spectra.  The model predicts the
number of images as a function of separation, shown in figure
\ref{fig:pos}.



The separation of images is related to the separation of folds.
For the majority of our simulation $\delta \sim 1$, and the separation is
given by the projection correlation length $\sigma_y$.  For $\delta \ll
1$, the angular density of folds becomes very rare, proportionate to an error
function.

The number of images is determined by the number of light folds along
the line of sight, i.e. how often the deflection angle is larger than
the separation to the line of sight in Figure \ref{fig:dt}.  Each
projected fold results in two fold singularities and thus four images.

\begin{figure}
\centerline{\epsfig{file=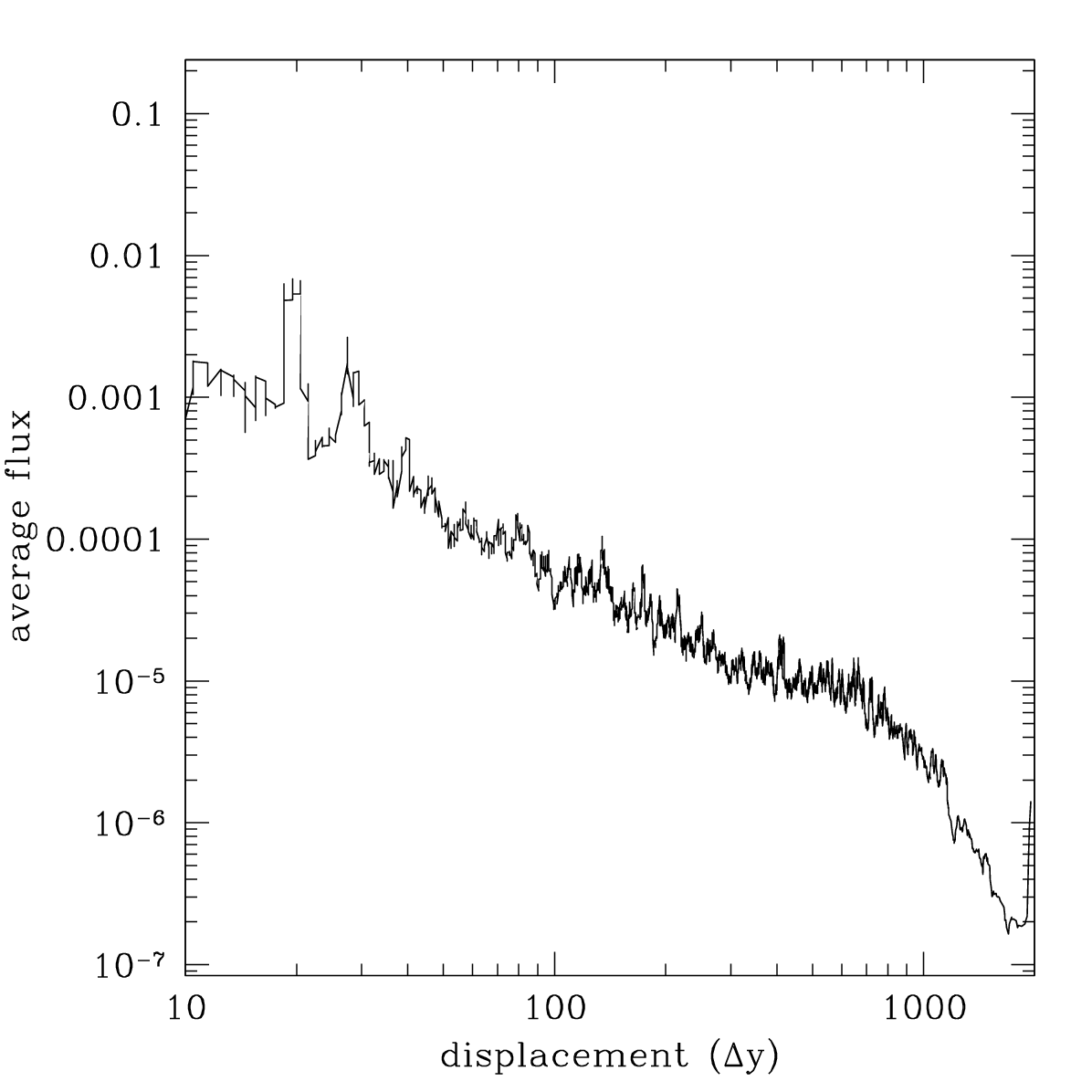,width=3in}}
\caption{Average Spatial Distribution of flux per logarithmic distance
  interval.  In a Gaussian model, the flux per image drops as
  $1/\theta$. 
}
\label{fig:posmag}
\end{figure}

The maximum deflection angle of an image is the change of refractive
index in the sheet, amplified by the alignment angle on a fold
singularity.  For a sheet much thinner than the projected fluctation
amplitude, the maximum projected density enhancement $C\equiv
\rho/\bar{\rho}$ relative to 
the projected mean sheet density $\bar{\rho}\sim n_e \Delta z/\alpha$
at a 
fold singularity
is the square root of the radius of curvature to the thickness of the
sheet $\tau$, $C=\sigma_y/\sqrt{A\tau}$
in the limit $\sigma^2/A \gg \tau$.


\section{Simulated Dynamic Spectra}

\begin{figure*}
\centerline{\hspace{0.32in}\includegraphics[width=3.15in]{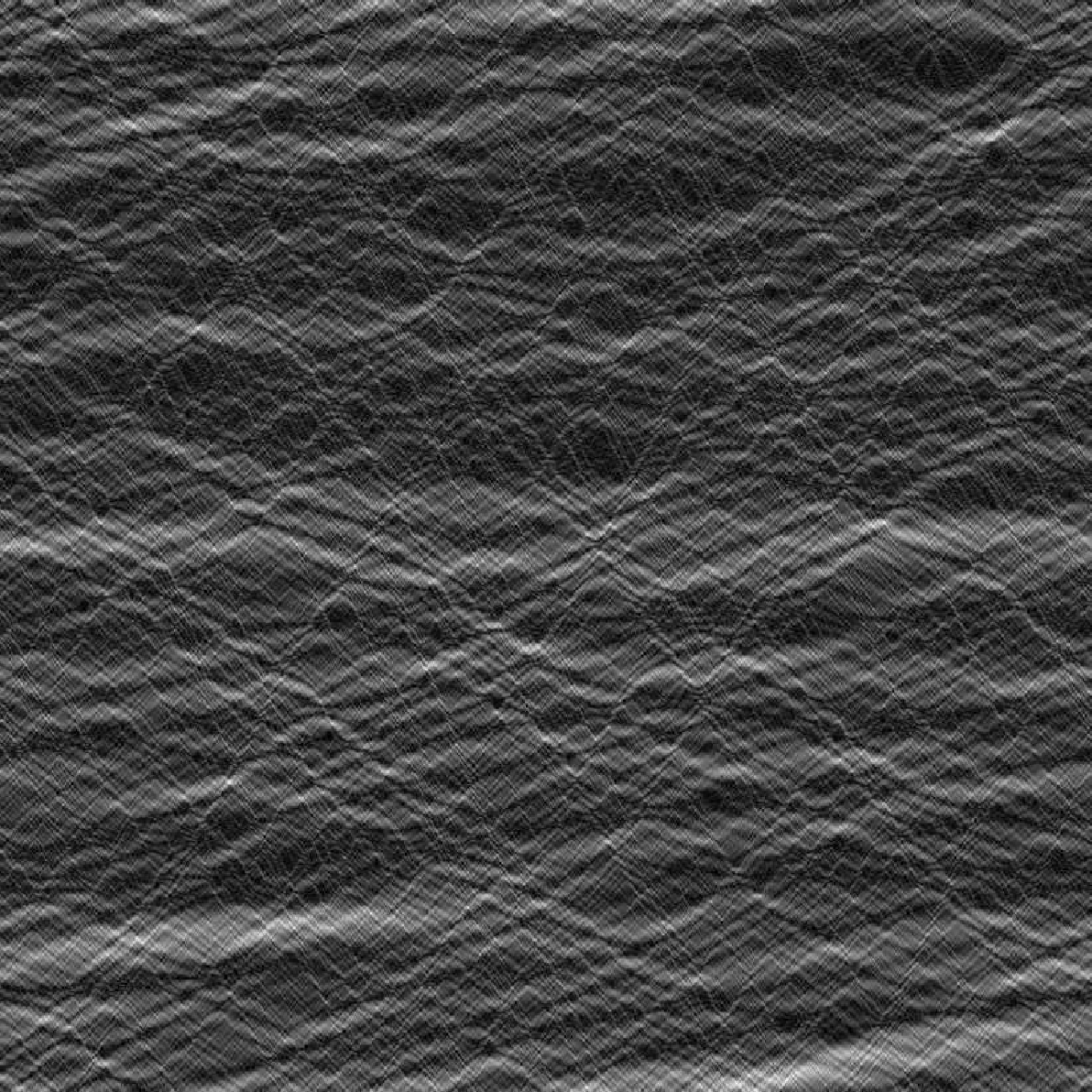}\includegraphics[width=3.15in]{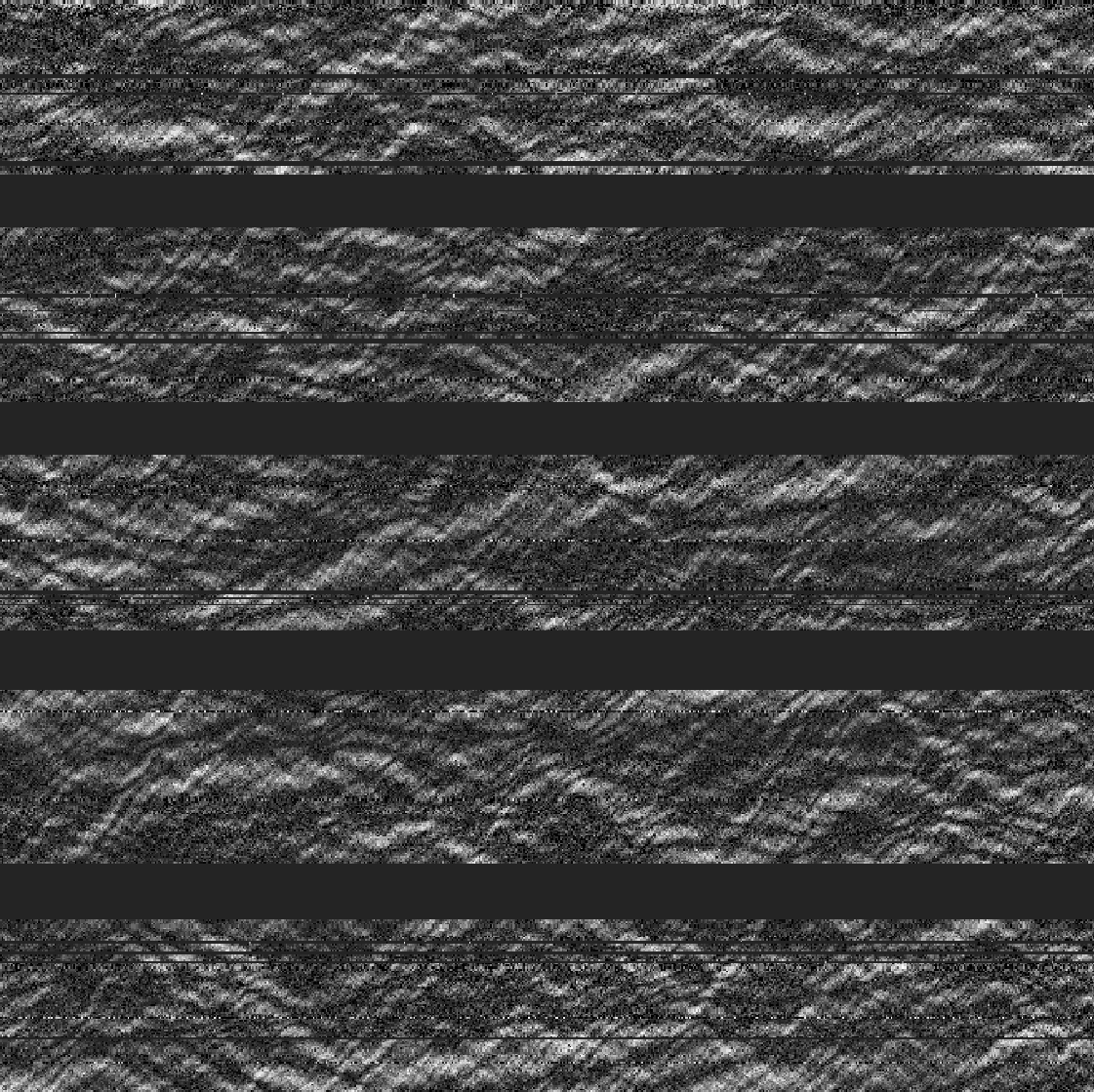}}
\vspace{-3.58in}
\centerline{\hspace{-0.079in}\epsfig{file=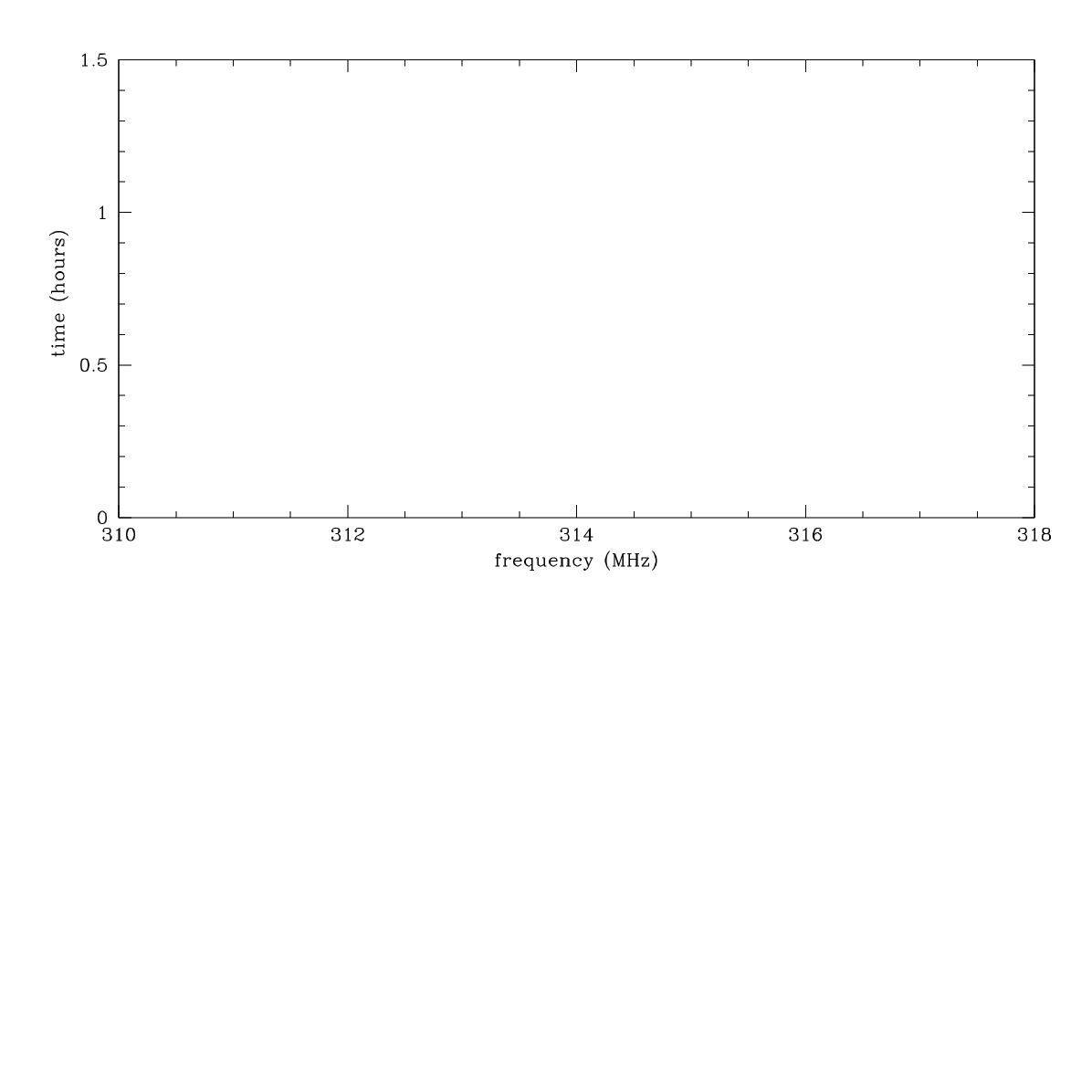,width=7.5in}}
\vspace{-3.58in}

\caption{dynamic pulsar spectrum.  Left panel is the simulated
  spectrum. Right panel is the dynamic spectrum of PSR B0834+06 from
  the Brisken et al (2010) dataset.
Horizontal axis is time, vertical
  axis is frequency.  For B0834, the horizontal axis is 4 MHz of
  bandwidth and the vertical axis 1.5 hours of time.
We reproduce the characteristic criss-cross
  pattern observed in real scintillation spectra.}
\label{fig:ds}
\end{figure*}

With the density field, we can solve the lens equations to simulate
dynamic spectra.  By adding the voltage field from each image  with their
appropriate amplitude and phases, we simulate the dynamic spectrum,
shown in figure \ref{fig:ds}.

\begin{figure*}
\centerline{\hspace{0.26in}\includegraphics[width=5.95in]{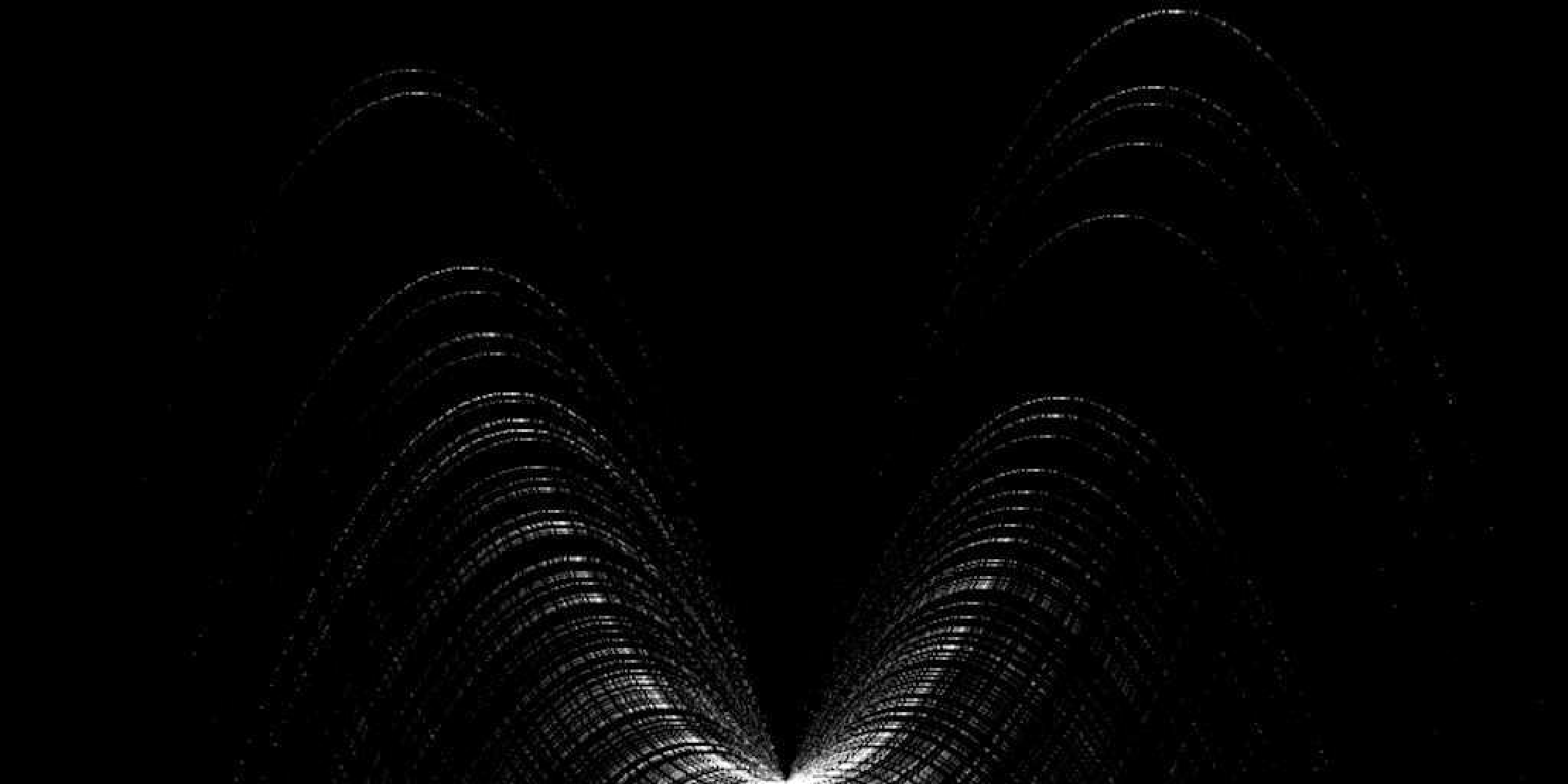}}
\vspace{-3.37in}
\centerline{\hspace{-0.159in}\epsfig{file=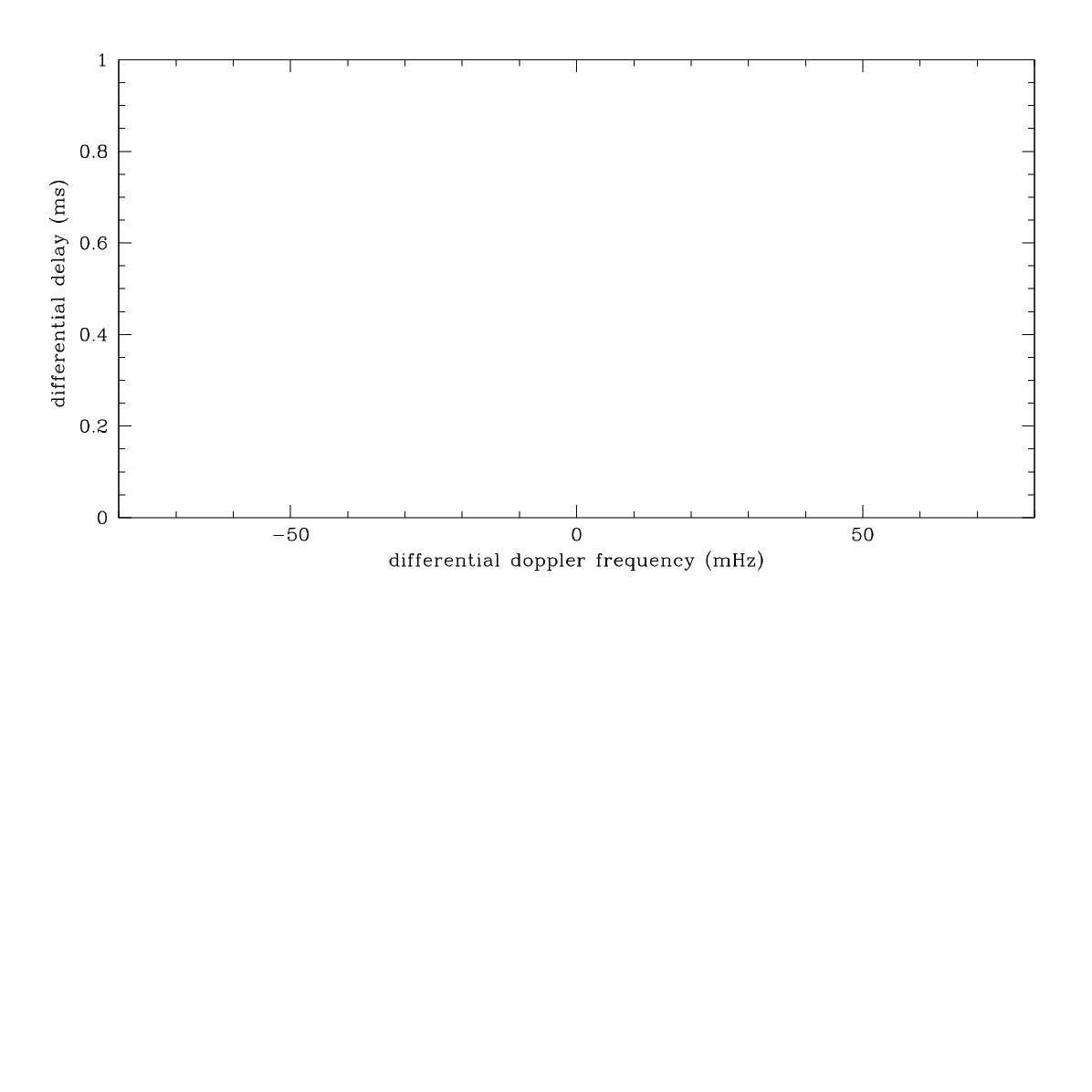,width=7.06in}}
\vspace{-3.37in}
\caption{simulated model secondary pulsar spectrum.  The inverted
  parabolic arcs at 
  large delay arise naturally in this model from the interference of
  the distant lensed images at the apparent sheet fold singularity with
  the brighter, closer in images.  The axis scales are meant to be
  illustrative, and depend on distance, pulsar velocity, etc}
\label{fig:ss}
\end{figure*}

A 2-D fourier transform maps this dynamic spectrum into a secondary
spectrum, shown in figure \ref{fig:ss}.

We find that the interference of these discrete, co-linear images
forms the inverted parabolic arcs, qualitatively similar to those that
are observed in \cite{2001ApJ...549L..97S} and
\cite{2005ApJ...619L.171H}.

\section{Discussion}

We can estimate the length scales involved in making the current sheet that
would produce the observed scintillation pattern.  This theory requires as
input a current sheet thickness, inclination angle, curvature,
amplitude of waves, and dissipation scale.

The thickness of the sheets can be estimated by considering the magnification of
images.  As shown in \cite{2012MNRAS.421L.132P}, the flux is roughly
the thickness divided by the impact parameter.  This follows from flux
conservation of lensing: the net flux is conserved, and flux changes
by order unity at impact parameters of order the physical size of the
lens, so the typical flux off-axis is roughly the ratio of the
furthest distance at which at image forms, to the size the lens.

The projected wavelengths
are observed to be of order $\sim 10$ AU;
this suggests a typical thickness of $h\lesssim 0.1$AU to explain the
$\sim$ 1\%  scattering intensities observed in \cite{2010ApJ...708..232B}.


The largest observed deflection angles at meter wavelength are $\gamma
\sim 0.01$", which requires an electron density change $\delta n_e\sim
100\alpha/C$ cm$^{-3}$.  For a typical interstellar plasma density of
$n_e \sim 0.03$ determined from pulsar dispersion, we need an
alignment of a fraction of a degree, with fluctuation amplitude $C\sim
30$ and a wavelength of $\lambda \sim 1000$ AU.  This combination
is not unique.


As discussed in \cite{2012MNRAS.421L.132P}, the  phenomenology of the
Extreme Scattering Events
prefers underdense lenses.  In an underdense sheet, the maximal change
of density is the density itself.  
With the assumption
$\delta n_e\sim n_e$, we obtain $\alpha \sim
10^{-2}$.  The probability of seeing a sheet at such an angle is $\sim
\alpha^2$, requiring the existence of $\sim 1/\alpha^2$ sheets along
the line of sight: if we imagine each ``sheet'' to be a thin disk, the
probability of seeing a disk at angle $\alpha$ gets one contribution
from the intrinsic alignment distribution, and one more from the
reduced geometric cross section of an aligned sheet.

For pulsar B0834+06, the distance is $\sim 0.64$ kpc (as determined
from the Dispersion Measure, and consistent with VLBI and ISM
geometries\citep{2010ApJ...708..232B}, also direct parallax: Deller and Brisken, unpublished 
), giving a typical sheet separation of
$s \sim 0.1$ pc.





These estimates are qualitative.  One expects current sheets to come
in a range of sizes, curvature and perturbation amplitude.  The
thickness might also vary.

\begin{figure*}
\centerline{\includegraphics[width=7.5in]{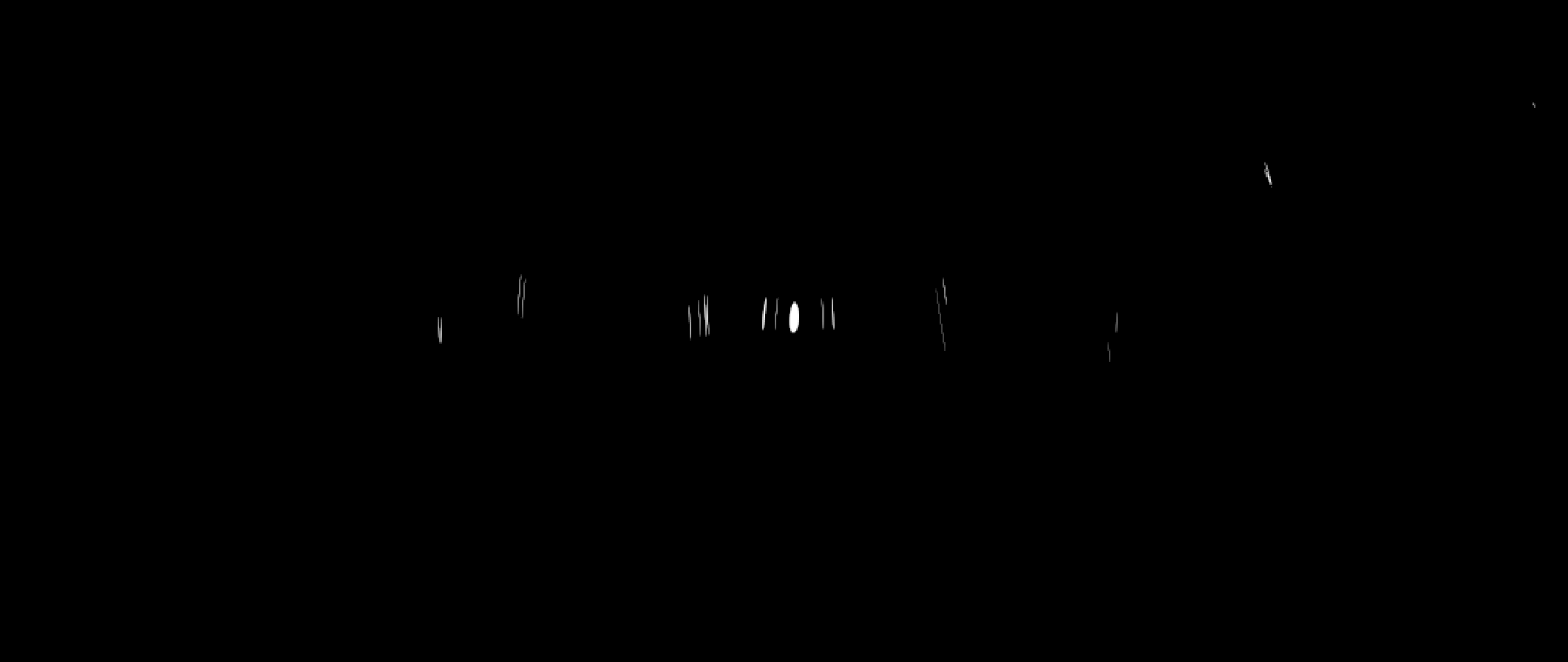}}
\caption{Simulated scattering image of the pulsar. A typical scale for this image would be 100 mas
  end-to-end.  The pulsar is
  modeled as a unit disk, with a much exagerated diameter of 2 mas. 
The actual
  size of the pulsar emission spot would be $\sim 100$  
  picoarcseconds. Flux
  conservation requires that the sum of image areas equals the
  original disk.  Thus, each pixel is either white or black in this figure.
This compares favourably with the VLBI reconstructed positions of 
\protect\cite{2010ApJ...708..232B}, where Fig 5 also shows images closely
aligned along a straight line. 
}
\label{fig:2d}
\end{figure*}

One of the attractive features of our mechanism is that it explains
very naturally the 1-d image of Brisken et al.~(2010). The reduced
dimensionality of the scattering image comes from the fact that the
deflection created by the screen is mostly in the direction
perpendicular to the screen's line of nodes. Therefore, the scattering
clumps will also form a line that is perpendicular to the line of
nodes. This is demonstrated in Fig. \ref{fig:2d}, where we show a simulated 2-d
scattering image of a pulsar. The alignment of the scattering clumps
is apparent.

\section{Future Potential}

In our picture, pulsar scintillation is dominated by a small number of
magnetic discontinuities highly aligned to the line of sight.  Surface
waves will move very slowly in this projected geometry, allowing for a
precise determination of the geometric properties.  This allows the
use of these sheets as lenses to study both pulsars and the
ISM\citep{2013arXiv1301.7505P}.  
It could also have application to
Intra-day variables\citep{2006A&A...446..185M} and the hyperstrong
scattering at the galactic center. 


\subsection{ISM dynamics}

Free parameters in our model include the thickness of the sheet and
the inclination angle.  These can be inferred by broad band
measurements of pulsar scintillation, as follows. Firstly, the
apparent position of images is expected to shift by a distance of
order of the sheet width, as one decreases the observing frequency
from the highest critical frequency at which the image forms, to a
factor of two below\citep{2012MNRAS.421L.132P}.  Secondly, the
co-linearity of the images is related to the inclination angle of the
sheet: the more aligned is the sheet with the line of sight, the
greater is the aspect ratio of the scattering image.

\subsection{Pulsar Emission imaging}

A straightforward application is the study of the reflex motion of the
emission region of the pulsar across the pulse phase.  One expects the
apparent emission region to move by distance of order of the effective
emission height multiplied by the ratio of the pulse width to the
rotation period.  Using VLBI mapping of the scattering geometry, one
can precisely predict the change of scintillation pattern as a
function of pulse phase.  The effective astrometric precision can be
sub nano arcsecond\citep{2013arXiv1301.7505P}.

\subsection{Distance Measurement}

It is tempting to use these lenses to obtain precision parallax
distances to pulsars.  It is difficult to keep the lens stable over a
year, when typical pulsars move by much more than an astronomical
unit, making the differential measurements challenging.  For pulsars
in binary systems, as is typical in millisecond pulsars, the orbital
motion about its companion will modulate the scintillation pattern.
This can be used to solve for the precision distance to the pulsar.

In the sheet lensing scenario, one can imagine obtaining widely
separated scattering images: the lensing angle scales as $\propto
\lambda^2$, so at low frequencies, for example with the LOFAR LBA,
hundreds of AU are probed.

This could result in direct parallax distances with nano arc
second precision, enough to determine pulsar distances for coherent
gravitational wave detection \citep{2012PhRvD..86l4028B}.

\section{Conclusions}

We have presented a quantitative theory of pulsar scintillation
inverse parabolic arcs.  These extend recent ideas of
\citet{2006ApJ...640L.159G} and \citet{2012MNRAS.421L.132P} about thin
current sheets as the scattering objects in the ISM, which naturally
explain the large angle scattering observed in pulsars and some
extragalactic sources.

This picture could explain all scintillation phenomena from refractive
lensing, all with structures greater than 0.1 AU.  The apparent
diffractive structure results from the interference between refractive
images, and no diffractive scattering is needed.

In this scenario, VLBI monitoring of pulsars on time scales of weeks,
at multiple low frequencies, can allow forecasts of the scattering
behaviour.  This in turn could improve gravitational wave timing residuals.
The same scenario also enables the coherent use of the scattered
images as a gigantic interstellar interferometer to map the motions of
the pulsar emission regions.

\section{Acknowledgements}

U-LP thanks NSERC and CAASTRO for support. YL is supported 
by the Australian Research Counsil Future Fellowship.

\newcommand{\araa}{ARA\&A}   
\newcommand{\afz}{Afz}       
\newcommand{\aj}{AJ}         
\newcommand{\azh}{AZh}       
\newcommand{\aaa}{A\&A}      
\newcommand{\aas}{A\&AS}     
\newcommand{\aar}{A\&AR}     
\newcommand{\apj}{ApJ}       
\newcommand{\apjs}{ApJS}     
\newcommand{\apjl}{ApJ}      
\newcommand{\apss}{Ap\&SS}   
\newcommand{\baas}{BAAS}     
\newcommand{\jaa}{JA\&A}     
\newcommand{\mnras}{MNRAS}   
\newcommand{\nat}{Nat}       
\newcommand{\pasj}{PASJ}     
\newcommand{\pasp}{PASP}     
\newcommand{\paspc}{PASPC}   
\newcommand{\qjras}{QJRAS}   
\newcommand{\sci}{Sci}       
\newcommand{\solphys}{Solar Physics}       %
\newcommand{\sova}{SvA}      
\newcommand{\aap}{A\&A}

\def\prd{{\it Phys.~Rev.~D\,}}

\bibliography{swave}
\bibliographystyle{mn2e}

\label{lastpage}

\end{document}